\documentclass[12pt]{article}
\usepackage{aaspp4}
\usepackage{epsf,astrobib,rotate,lscape}
\usepackage{times,mathptm}

\def\etal{{\it et al.}}
\def\simlt{\hbox{ \rlap{\raise 0.425ex\hbox{$<$}}\lower 0.65ex\hbox{$\sim$} }}
\def\simgt{\hbox{ \rlap{\raise 0.425ex\hbox{$>$}}\lower 0.65ex\hbox{$\sim$} }}
\def\lta{\mathrel{\spose{\lower 3pt\hbox{$\sim$}}\raise 2.0pt\hbox{$<$}}}
\def\gta{\mathrel{\spose{\lower 3pt\hbox{$\sim$}}\raise 2.0pt\hbox{$>$}}}

\def\psd{deg$^{-2}$}

\lefthead{Weinberg \& Nikolaev}
\righthead{LMC Structure}

\begin{document}

\title{Structure of the Large Magellanic Cloud from 2MASS}
\author{Martin D. Weinberg \& Sergei Nikolaev}
\affil{Department of Physics \& Astronomy\\%
  University of Massachusetts, Amherst MA 01003-4525}

\begin{abstract}
  
We derive structural parameters and evidence for extended tidal debris
from star count and preliminary standard candle analyses of the Large
Magellanic Cloud based on Two Micron All Sky Survey (2MASS)
data.  The full-sky coverage and low extinction in $K_s$ presents an
ideal sample for structural analysis of the LMC.

The star count surface densities and deprojected inclination for both
young and older populations are consistent with previous work.  We use
the full areal coverage and large LMC diameter to Galactrocentric
distance ratio to infer the same value for the disk inclination based
on perspective.

A standard candle analysis based on a sample of carbon long-period
variables (LPV) in a narrow color range, $1.6<J-K_s<1.7$ allows us to
probe the three-dimensional structure of the LMC along the line of
sight.  The intrinsic brightness distribution of carbon LPVs in
selected fields implies that $\sigma_M\simlt 0.2^m$ for this color
cut.  The sample provides a {\it direct} determination of the LMC disk
inclination: $42.3^\circ\pm 7.2^\circ$.  

Distinct features in the
photometric distribution suggest several distinct populations.  We
interpret this as the presence of an extended stellar component of the
LMC, which may be as thick as $14$~kpc, and intervening tidal debris
at roughly 15 kpc from the LMC.

\end{abstract}


\newpage


\section{Introduction} \label{sec:intro}

Morphologically, the LMC is an irregular barred spiral galaxy with
three spiral arms and an extended outer loop of stellar material
\cite{dvc73}.  Based on deprojection and photometric distances 
(e.g.  de Vaucouleurs 1957, 1980), its disk is inclined at an 
angle of $27^\circ$ to the plane of the sky.
The disk exhibits solid body rotation out to $2.5^\circ$ with a
rotation center at $5^h21^m, -69^\circ 17'\, (1950)$, about
$0.6^\circ$ north of the optical center of the bar.  This kinematic
signature is present in a variety of tracers: HI gas, planetary
nebulae, HII regions, supergiants, CH stars, etc.  Freeman \etal{}
(1983) have examined kinematics of rich star clusters with ages
between $100$ Myr and $10$ Gyr.  They found that young clusters
rotated with HI gas, while the older ones (SWB VII; Searle \etal{}
1980) formed a flattened rotating system with dispersion along the
line of sight $\sigma \sim 18$ km/s.  A later study of more extended
sample of outer LMC clusters \cite{sch92} confirmed the absence of
isothermal pressure-supported spheroid.  All this has led to the
standard view that the LMC is a {\em geometrically thin} object.

However, recent studies have suggested that the LMC may have an
extended component.  First, the evidence for a flattened spheroid
population was found in the kinematics of old long-period variables
\cite{hug91}.  Kunkel \etal{} (1997) describe a population of carbon
stars out to 12 kpc from the LMC center.  These authors interpret
these in the context of a thin disk model and derive a rotation curve
and mass estimate.  However, Weinberg (2000) argues that the LMC
should be evolving rapidly in the Milky Way tidal field, based on both
analytic calculations and n-body simulations.  The tidal field causes
the LMC disk axis to precess and torques disk orbits out of the disk
plane, causing a strongly flared, spheroidal-like distribution in the
outer Cloud and loss of stars and gas.  This interaction leads to an a
spatially extended population while roughly preserving the disk-like
kinematic signature (i.e., small $\sigma$).

The detection of, or strict limits on, the predicted extended
distribution would resolve these views and is one of the goals for the
present study.  Our star count analysis is based on fitting the
projected spatial density of several LMC populations among those
identified in Nikolaev \&~Weinberg (2000; hereafter NW00) based on
their location in the color-magnitude diagram (CMD) of the field.  The
late-type giant populations are dominated by the LMC and may be used
as tracers of the spatial structure of the Cloud.  Each population is
fitted by two models: 1)~thin exponential disk and 2)~spherical power
law model.  Our best-fit models based on circular disks give the
inclination of the LMC to the line of sight $i \approx
24^\circ-28^\circ$ and the position angle of the disk $\theta \approx
165^\circ-174^\circ$, in good agreement with previous estimates.  The
direction of the LMC disk inclination is also determined.  In short,
projected 2MASS star counts reproduce the standard LMC inferences.

The near-infrared 2MASS photometry easily discriminates carbon stars
in the color-magnitude diagram.  While the 2MASS single-epoch survey
of the LMC does not provide variability information, we identified a
region of the CMD populated nearly exclusively by carbon-rich AGB
stars (Region~J in NW00), which are also long-period variables (LPVs).
This identification is reinforced by recent analyses of MACHO data
\cite{alv98,woo99}, although roughly $25\%$ could be binaries
\cite{woo99}.  LPVs obey period-luminosity-color (PLC) relations
(e.g., Feast \etal{} 1989), and based on the PLC relations, the LPVs
in a narrow color range ($1.6<J-K_S<1.7$) are standard candles with
$\sigma_K\approx0.3^m$.  Their photometric distribution in selected
LMC fields has a least three distinct components: a well-defined
narrow distribution due to LMC disk and two secondary peaks at fainter
and brighter magnitudes.  The differential photometric distance to the
central disk peak provides a direct determination of the inclination:
$42.3^\circ\pm7.2^\circ$.  This value is consistent with but larger
than those inferred by deprojecting isopleths.  The secondary peak
could in principle be due to stellar blends, geometric structure,
interstellar reddening, distribution of periods, gradient in age and
metallicity, or contaminating population of overtone pulsators in the
sample.  We carefully examine the possible origins of this secondary
component, and conclude that most likely explanation is spatial
structure.  This interpretation is bolstered by the good match of the
central peak, which includes known fundamental mode pulsators, with
the established LMC disk inclination.  This implies the presence of an
extended component of the LMC, which may be as thick as 14~kpc along
the line of sight.  This component is thicker than the 2.8 kpc
flattened spheroid suggested by Hughes \etal{} (1991) from kinematic
data and may be streams of material rather than be smooth and
well-mixed.  The bright peak is then an intervening population at a
distance of roughly 35 kpc and the existence of relatively young
carbon-rich AGB stars suggests tidal debris.

The outline for the paper follows.  In the main section, we briefly
describe the 2MASS sample (\S\ref{sec:data}), and present the star
count (\S\ref{sec:ml}) and the standard candle (\S\ref{sec:sca})
analyses of the LMC.  In particular, \S\ref{sec:sca_select} describes
the selection of standard candles, \S\ref{sec:sca_method} gives the
details of the standard candle analysis and \S\ref{sec:sca_results}
presents the main results.  The major conclusions are summarized in
\S\ref{sec:summary}.

\section{Observations} \label{sec:data}

\begin{figure}[thb]
\mbox{
\mbox{\epsfysize=8.2cm\epsfbox{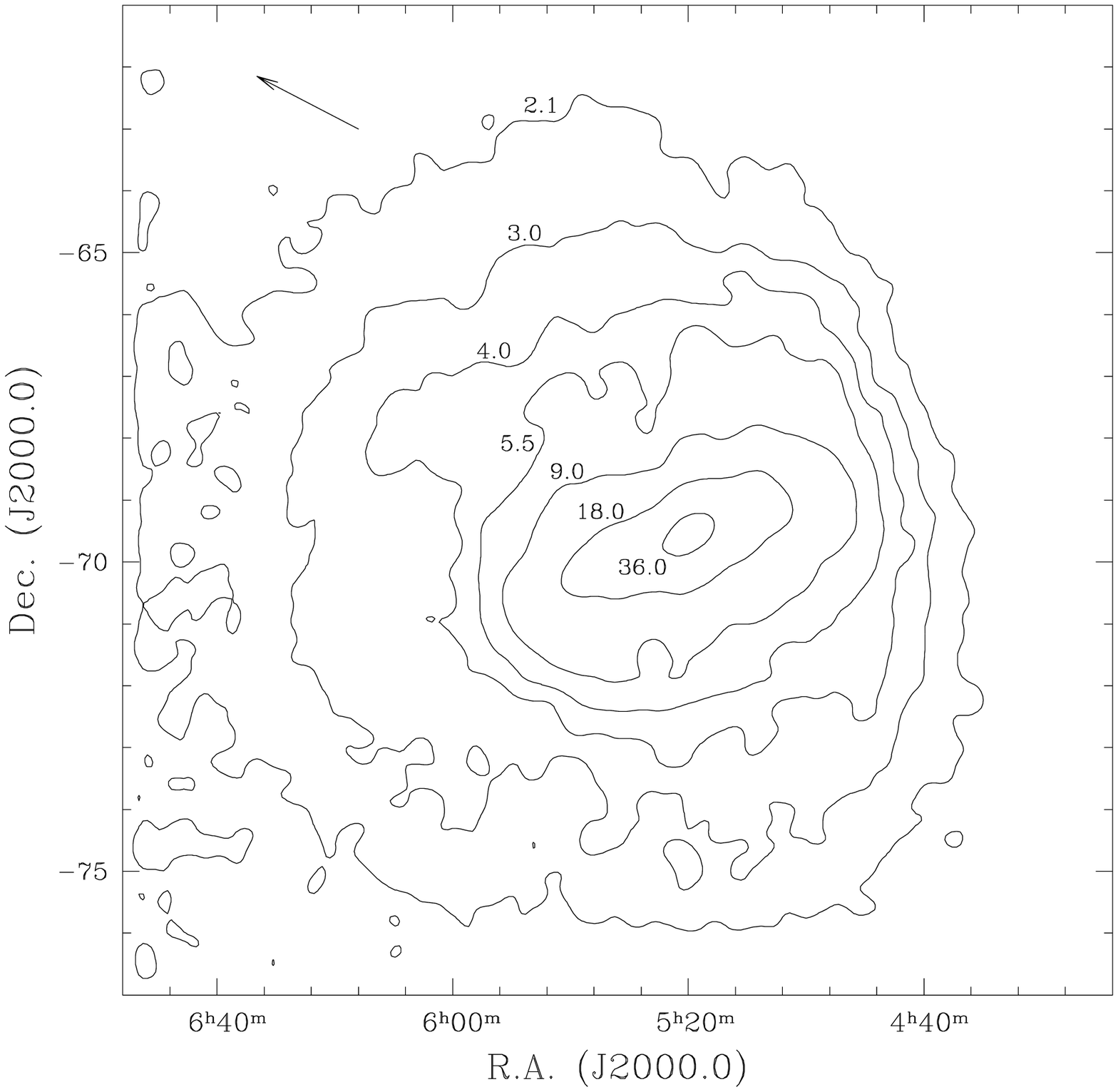}}
\mbox{\epsfysize=8.2cm\epsfbox{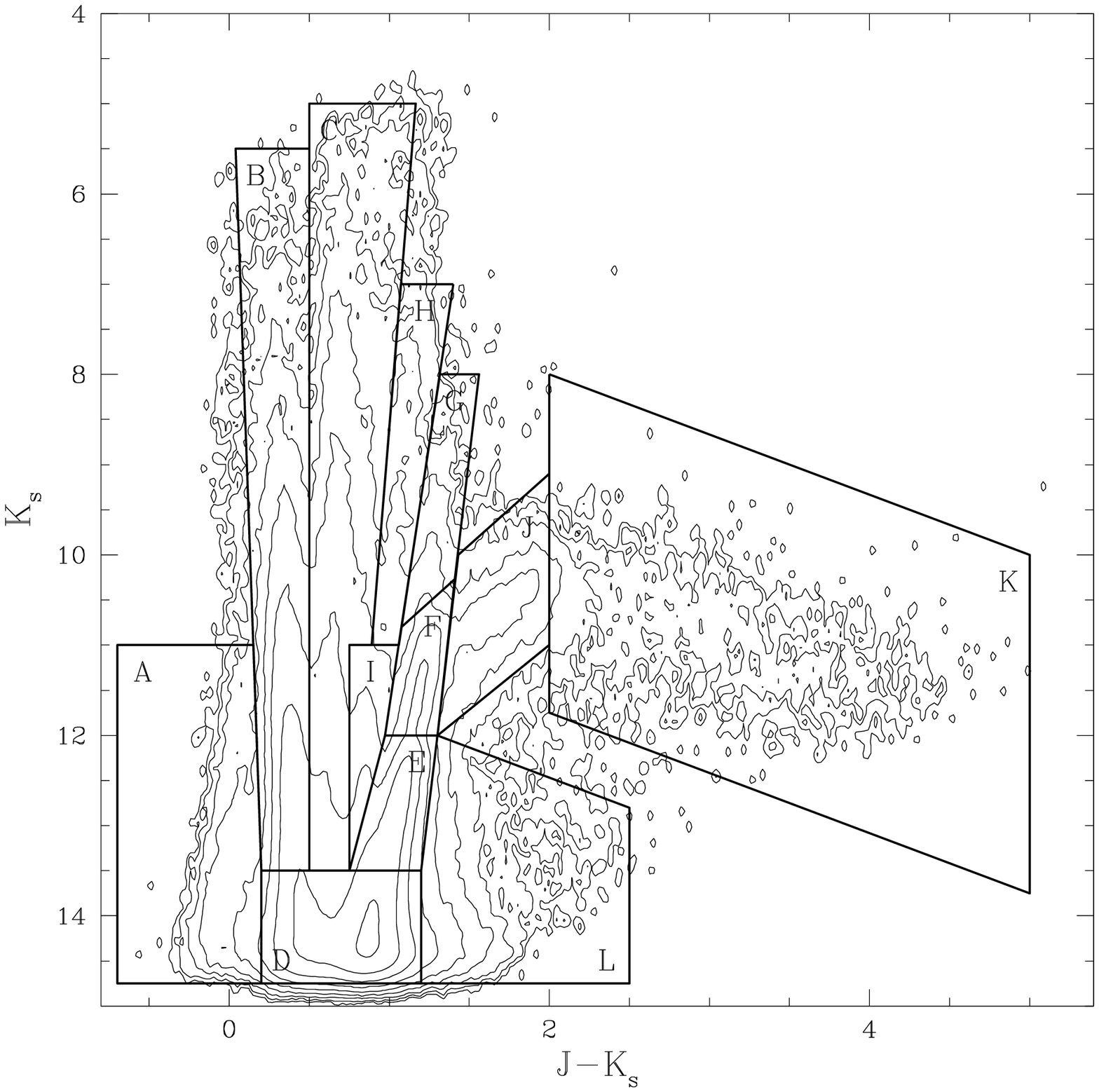}}
}
\caption{{\em Left panel}: Density distribution of 2MASS sources
  in the LMC field.  Contour levels are labeled in units of $10^3$
  \psd{}.  The arrow points in the direction of Galactic center.  
  {\em Right panel}: CMD of the field showing 12 regions 
  corresponding to major features of the diagram.  Contour levels 
  are logarithmic, from $2.0$ to $6.0$, spaced by $0.5$.  
  \label{fig1}
}
\end{figure}

The LMC field, (4$^h$00$^m$ to 6$^h$ 56$^m$ in right ascension,
$-78^\circ$ to $-60^\circ$ in declination, J2000.0) has been observed
by 2MASS and is included in the most recent data release.  Details of
the data reduction and sample selection are described in NW00.
Figure~\ref{fig1} shows both the projected spatial distribution of
sources and the color-magnitude diagram of the field.
The color-magnitude diagram also shows the location of 12 regions
analyzed in NW00.  Total number of sources in our sample is 
823,037.  

NW00 examined the populations in selected CMD regions and associated
the features with known populations of stars (cf. Table~2 of NW00).
Here, we take an in-depth look at the spatial distribution of sources
in seven of those regions (Table~\ref{regions}) which contain mostly
LMC stars.  These seven areas of the CMD account for a quarter of all
sources in the field.

\begin{table}[h]
\centering
\caption{Stellar populations of selected CMD regions.  
\label{regions}}
\begin{tabular}{cccl}
\hline
Region & $N_{src}$ & ${f_{Gal}}^{\rm a}$ & LMC populations \\
\hline
A &   6,659 & 0.15 & Young O,B,A supergiants, O3---O6 dwarfs \\
E & 166,263 & 0.05 & Low- and intermediate-mass RGB stars, E-AGB stars \\
F &  22,134 & 0    & Oxygen-rich AGB stars, E-AGB and TP-AGB, LPVs \\
G &   1,438 & 0    & Luminous E-AGB stars, O-rich LPVs \\
H &   2,450 & 0.05 & Red supergiants, luminous E-AGB stars \\
J &   8,229 & 0    & Carbon-rich TP-AGB, LPVs \\
K &   2,212 & 0    & Dust-enshrouded C-rich TP-AGB, OH/IR, cocoon stars \\
\hline
Total & 209,385 \\
\hline
\end{tabular}
\vskip 0.1cm

$^{\rm a}$Fraction of Galactic sources estimated from synthetic 
model (see NW00)
\end{table}

\begin{figure*}
\epsfysize=18.0cm
\centerline{\epsfbox{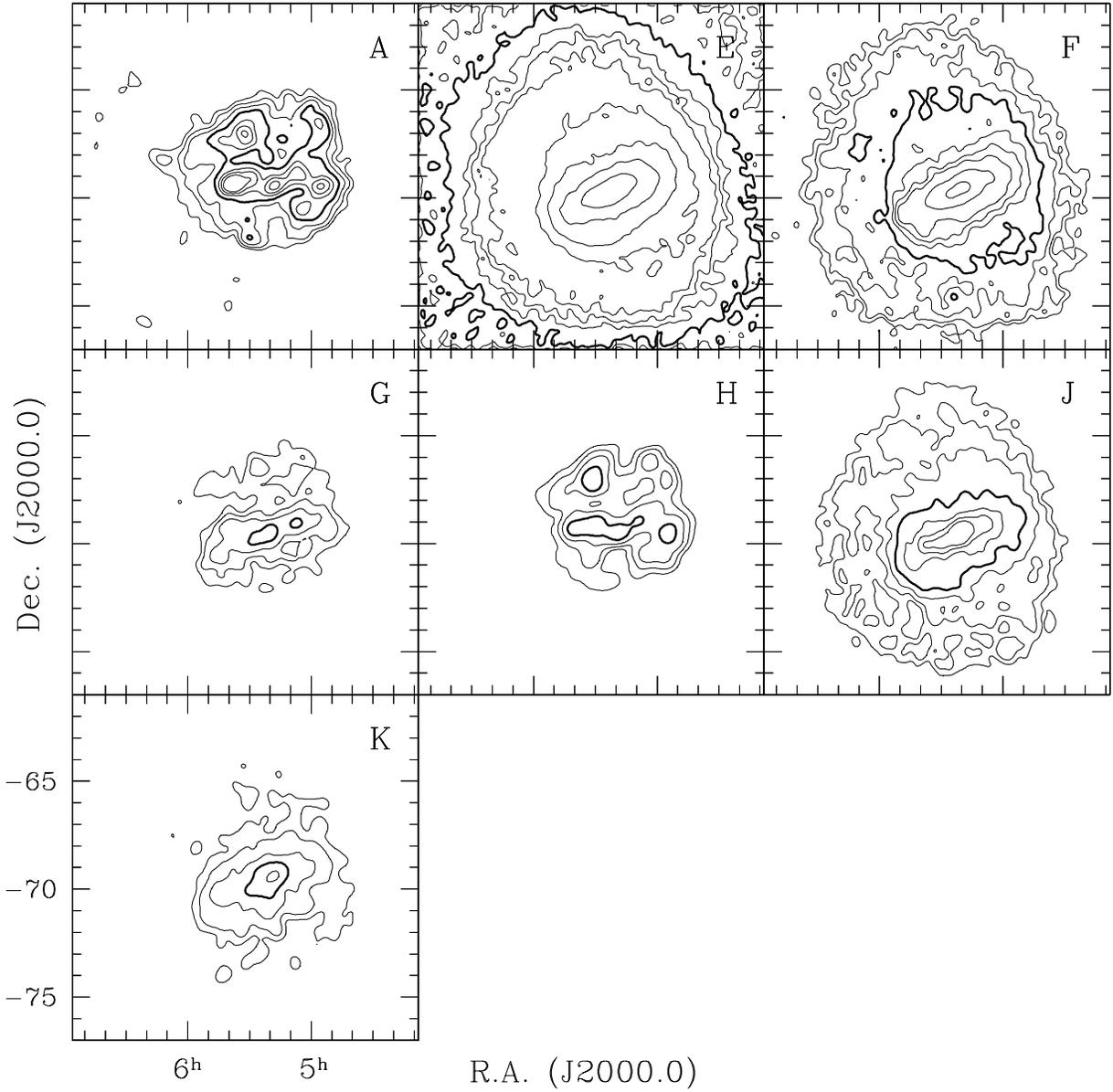}}
\caption{Projected spatial density distributions of 2MASS sources 
  in CMD regions.  The contour levels are 15, 30, 60, 120, 240, 
  350, 480, 960, 1920, 3840, 7680 \psd{}.  In panel E the lowest 
  contour level is 60 \psd{}.  The contour level of 120 \psd{} is 
  highlighted for convenience.  \label{fig2}}
\end{figure*}

The projected spatial density distribution for each of the seven
regions is shown in Figure~\ref{fig2}.  As expected, younger 
populations (Regions A and H) have relatively clumpy distributions
which trace the spiral pattern of the LMC (cf. Figure~7 of 
Schmidt-Kaler 1977).  Older stars, on the other hand, have 
smoother and more extended distributions with significant 
overdensity in the bar of the Cloud.  Several well-known 
morphological features of the LMC are easily recognizable in 
Figure~\ref{fig2}, e.g. 30~Doradus complex (an HII region near 
$\alpha=5^h36^m$, $\delta=-69^\circ$) and asymmetric outer loop 
in the south-eastern part of the LMC, traced by AGB stars.

\section{Spatial Structure Using Parametric Maximum Likelihood} 
  \label{sec:ml}

To quantify the spatial distribution of sources in six selected
regions, we perform maximum likelihood (ML) analyses for thin
exponential disk and spherical power-law models.  The observed 
source counts for each population are binned in equatorial 
coordinates, $n_{ij}^o,\,\,i=1,N;\,j=1,M$.  The scale was chosen 
by successively reducing the bin size until the inferred 
parameters converged.  The final coordinate grid is spaced 
uniformly, with the bin size $\Delta\alpha=\Delta\delta=1^\circ$.
The ML scheme selects a parametric model for which the expected 
source counts $n_{ij}^e$ most closely match the observed source 
counts $n_{ij}^o$.  The goodness-of-fit measure,
\[
u^2 = \sum _{i,j} \frac {(n_{ij}^o-n_{ij}^e)^2}{n_{ij}^e},
\]
is distributed as $\chi^2$ with $(N \times M-1-n_p)$ degrees of
freedom, where $n_p$ is the number of free parameters of the model
(see below).  The expected source counts $n_{ij}^e$ are obtained 
from the corresponding source density in each bin, $\rho_{ij}^e$,
predicted by the model:
\[
n_{ij}^e = \frac {N_{src}\, \rho_{ij}^e}{\sum _{i,j} \rho_{ij}^e}
\]
(see Appendix~{\ref{sec:appdetails}} for details).

The density of the exponential disk is described by
\begin{equation}
\rho = C\, {\rm exp} \, (-r/R), \label{eq:expd}
\end{equation}
where $R$ is the disk scale length.  The model has five free
parameters: coordinates of the disk center $\alpha_0$, $\delta_0$,
the scale length $R$, and two orientation angles: inclination $i$
($0^\circ \le i < 180^\circ$) and position angle (line of nodes)
$\theta$ ($0^\circ \le \theta < 360^\circ$).  The position angle
$\theta$ increases counterclockwise from the north (i.e. NESW).  
The
density of the spherical power law model is described by
\begin{equation}
\rho = S r^{-\nu}. \label{eq:powl}
\end{equation}
The free parameters are positions of the LMC center, $\alpha_0$ 
and $\delta_0$, and power law index $\nu$.

The results of parametric fits are presented in 
Table~\ref{table:expd} (for thin exponential disk model) and 
Table~\ref{table:powl} (for spherical power law model).  The best 
fit parameters in both tables can be grouped by the population 
age: results for young stars (A, H) and older stars (E, F, J, K) 
show good agreement among themselves.  The error in centroid of
each population is $\simlt 0.1^\circ$.  The near side position 
angles marked with `?' are inaccurate due to clumpiness and 
compactness of the corresponding distributions.

\begin{table*}[h!]
\centering
\caption{Best fit parameters of exponential disk model.  
\label{table:expd}}
\begin{tabular}{ccccccc}
\hline
Region&\multicolumn{6}{c}{Parameters}\\
&$\alpha_0 (^\circ)$&$\delta_0 (^\circ)$&$R$ (kpc)&$i (^\circ)$&
$\theta (^\circ)$&PA$_{near} (^\circ)$\\
\hline
A &80.9&$-68.9$&$1.63\pm0.01$&$35.8\pm0.8$&$86.6\pm1.5$&$-3?$ \\
E &80.7&$-69.6$&$1.59\pm0.01$&$24.0\pm0.4$&$171.0\pm1.0$&$81$ \\
F &80.6&$-69.5$&$1.36\pm0.01$&$26.5\pm0.9$&$171.4\pm2.2$&$81$ \\
G &80.8&$-69.3$&$1.32\pm0.03$&$30.9\pm2.4$&$116.5\pm5.1$&$27?$ \\
H &80.4&$-68.8$&$1.84\pm0.03$&$37.6\pm1.3$&$64.0\pm2.5$&$154?$ \\
J &80.8&$-69.5$&$1.37\pm0.01$&$25.2\pm1.5$&$168.6\pm3.9$&$79$ \\
K &80.5&$-69.5$&$1.40\pm0.02$&$27.8\pm2.4$&$173.2\pm5.5$&$83$ \\
\hline
\end{tabular}
\par\parbox[t]{14cm}{%
\vskip 0.1cm
The positional accuracy of distribution centroids ($\alpha_0$, 
$\delta_0$) is $\simlt 0.1^\circ$.  `?' indicates the large 
uncertainty in the position angle.  Parameters of the fit for 
young populations (A, G, H) carry large systematic error due to 
poor quality of the fit.}
\end{table*}

The distributions of young OB stars and supergiants are clumpy and
therefore poorly fitted by a smooth model.  The centroid of these
populations is $\sim 1^\circ$ to the north of the optical center 
of the bar (defined by the center of symmetry of the bar, at
$\alpha_{1950}=81.0^\circ$, $\delta_{1950} =-69.8^\circ$), similar
to the displacement found by de~Vaucouleurs~\& Freeman (1973).  
The scale lengths $R$ derived for these populations are noticeably
greater than the scale lengths for older populations and reflect 
the location of the distinct star forming activity.  The position
and inclination angles for these populations are mutually 
consistent.  Our derived inclinations, $i\approx36^\circ-
38^\circ$, are consistent with $i=38.2^\circ$ found from the 
distribution of HI regions (Feitzinger~\etal{} 1977), and also
with $i=36^{+2}_{-5}$~degrees found from Monte Carlo simulations
of ultraviolet photopolarimetric maps of the western LMC 
\cite{col99}.

The older populations (M giants, AGB stars and LPVs) are
well-represented by a smooth density law.  The centroids for these
populations are within $0.4^\circ$ of each other on the sky and 
are close to the optical center of the bar.  The scale lengths are
$R\approx 1.3-1.4$~kpc (sources in Region~E have the largest scale
length among older stars, $R\approx1.6$, which is probably due to
Galactic M dwarfs contamination).  The inferred inclinations are 
in the range $i\sim24^\circ-28^\circ$, in good agreement with
previous determinations from star counts \cite{dvc55}, 
$i=(25\pm5)^\circ$; distribution of star clusters, 
$i=(25\pm9)^\circ$ or HI isophotes, $i=(27\pm5)^\circ$ 
\cite{mcg66};
or photographic R isophotes \cite{dvc57}, $i=(27\pm2)^\circ$.  
These
deprojection-based position angles for the group are mutually 
consistent, $\theta\sim 169^\circ-173^\circ$, and fall in the 
range $\theta=160^\circ-180^\circ$ derived from surface photometry
of the LMC by others (see, e.g. Table~2 of Schmidt-Kaler~\& 
Gochermann 1992).

\begin{figure}[thb]
\epsfysize=9.0cm \centerline{\epsfbox{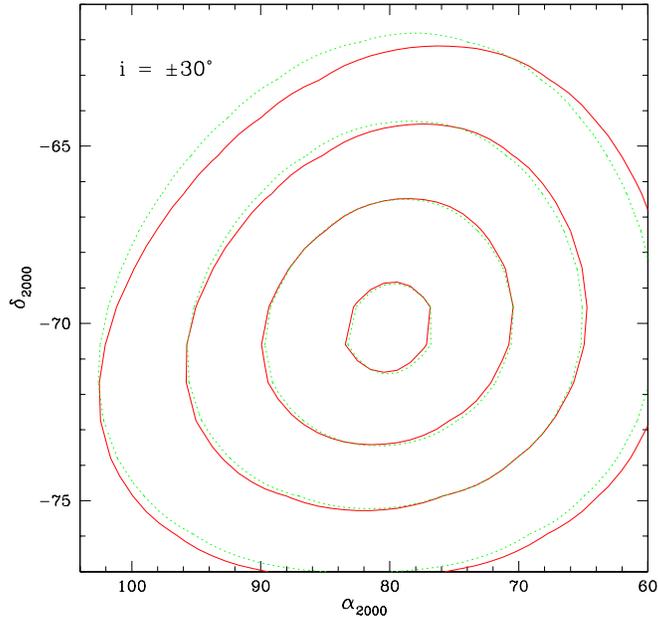}}
\caption{Isodensity contours of test models showing the 
  sensitivity to the direction of inclination.  Each test model is
  represented by 3000 sources distributed according to an 
  exponential disk with $\left\{\alpha_0,\delta_0,R,i,\theta 
  \right\}$ = $\left\{80^\circ,-70^\circ,1.5,\pm30^\circ, 
  135^\circ\right\}$.  Four logarithmically spaced density levels
  of 0.0003, 0.001, 0.003, 0.01 \psd{} are shown.  Solid line 
  shows model with $i=-30^\circ$, dotted line --- model with 
  $i=+30^\circ$.  The difference is clearly seen, especially in 
  outer contours.
\label{fig3}}
\end{figure}

The extended spatial coverage of the LMC field by 2MASS allows one to
determine the absolute direction of the inclination, i.e. to determine
the closest side of the LMC.  To illustrate this point, we make two
different test models of an exponential disk with
$\left\{\alpha_0,\delta_0,R,i,\theta \right\}$ $=$
$\left\{80^\circ,-70^\circ,1.5,\pm30^\circ,135^\circ\right\}$.  Both
disks are modeled with 3,000 point sources.  The restored density
contours are presented in Figure~\ref{fig3}.  The difference in the
expected source counts is clearly seen in the outer regions.  This
suggests that for relatively spatially extended populations both the
absolute value and the direction of the inclination can be reliably
determined.  On the other hand, if a population is relatively compact
in the sky, the inferred direction of inclination may differ from the
actual value (cf.  Table~\ref{table:expd}).  Based on the results for
older populations, we see that the nearest side of the LMC is its
eastern side, in agreement with previous results based on photometry
of Cepheids in the Cloud \cite{dvc55,gas78,lan86}.  Restricting our
attention to Region~J, which has little if any Galactic contamination,
we can be sure that we are not affected by a gradient caused by the
Galactic foreground.

The results of fits to a spherical power-law profile are described
in Table~\ref{table:powl}.  There are no significant difference 
between power-law exponents for various populations: all values 
are $\approx 2.5-2.6$.  The centroid shift for younger 
populations, present in Table~\ref{table:expd}, is seen here as 
well.  The underlying distribution, a disk and spheroid together,
is described here as a single profile.  A power-law disk profile,
of course, will have an index $1+\nu$, where $\nu$ is the index 
for the spherical profile, and therefore some unknown combination
of a disk and spheroid makes interpretation difficult.  We note 
that these fits have poorer quality than the exponential fits and
probably do not bear on the reality of a spheroidal population.  
A more independent test for spatially extended population is 
described below (\S\ref{sec:sca_method}).
\begin{table}[htb]
\centering
\caption{Best fit parameters of power law model.
\label{table:powl}}
\begin{tabular}{cccc}
\hline
Region&\multicolumn{3}{c}{Parameters}\\
      &$\alpha_0 (^\circ)$&$\delta_0 (^\circ)$&$\nu$\\
\hline
A & 80.8 & -69.0 & $2.60\pm0.01$\\
E & 80.5 & -69.9 & $2.52\pm0.01$\\
F & 80.3 & -69.9 & $2.59\pm0.01$\\
G & 80.7 & -69.1 & $2.65\pm0.02$\\
H & 80.2 & -69.0 & $2.55\pm0.02$\\
J & 80.7 & -69.9 & $2.56\pm0.01$\\
K & 80.4 & -69.9 & $2.59\pm0.02$\\
\hline
\end{tabular}
\vskip 0.1cm
The positional accuracy of distribution centroids ($\alpha_0$, 
$\delta_0$) is $\simlt 0.1^\circ$.
\end{table}

We compared the results of our fits to similar fits by Hughes~\etal{}
(1991), who modeled the distribution of intermediate and old
long-period variables (ILPV and OLPV, respectively) with exponential
disk and power law models.  They found $\nu=1.8\pm 0.1$ and
$R=1.6\pm0.2$ (OLPV), $\nu=1.7\pm0.1$ and $R=1.7\pm0.2$ (ILPV).  Our
results for similar populations (Regions E, F, G and J) imply
$\nu\sim2.5$ and $R\sim1.4$ kpc.  While the scale lengths $R$ are
mutually consistent, the best-fit power law exponents differ.  Our
best fit parameters are larger than the power-law model exponents of
Hughes~\etal{} which has $\nu\approx2.0$.

To summarize, the projected distribution of LMC populations 
observed by 2MASS is consistent with previous studies.  We found 
the scale length of the LMC disk, $R\sim1.4-1.5$ kpc, the 
inclination angle, $i\sim24^\circ-28^\circ$, and the direction of
the LMC tilt in good agreement with existing estimates.

\section{Standard Candle Analysis} \label{sec:sca}

We complement our previous analysis (\S\ref{sec:ml}) by 
incorporating photometric distances in addition to our CMD 
selection.  Below, we describe the selection of standard candles,
the details of the analysis, and the implications for the 
structure of the LMC derived from 2MASS photometry.

\subsection{Selecting Standard Candles from 2MASS Data} 
\label{sec:sca_select}
  
Good standard candles satisfy three conditions: 1)~they must be
luminous and easily identified; 2)~they must be sufficiently numerous
and representative of the underlying structure, and 3)~they must have
small photometric dispersion as a class (that is, small $\sigma_M$).
In NW00, we argued that stars in Region~J of the 2MASS CMD are
potentially good standard candles.  Being brighter and redder than the
RGB tip, most of these stars are carbon-rich thermally-pulsating AGB
stars (TP-AGB).  Recent data \cite{alv98,woo99} suggests that most of
these stars exhibit Mira-like variability.  The fraction of variables
in this region is close to $100\%$, although roughly $25\%$ could be
binaries \cite{woo99}.  Here, {\em we will assume that all sources in
  Region~J are carbon-rich long-period variables.}  Their red colors
effectively discriminate against the population of oxygen-rich LPVs,
since the latter rarely have $J-K_s>1.5$ \cite{hug90}.  As long-period
variables, these stars follow a linear period-luminosity-color
relation (e.g. Feast \etal{} 1989).  The luminosity of these stars can
be characterized by their periods or near-infrared colors, and
therefore, these stars can be used to probe the structure of the LMC
along the line of sight.

\begin{figure*}[p]
\epsfysize=17.0cm
\centerline{\epsfbox{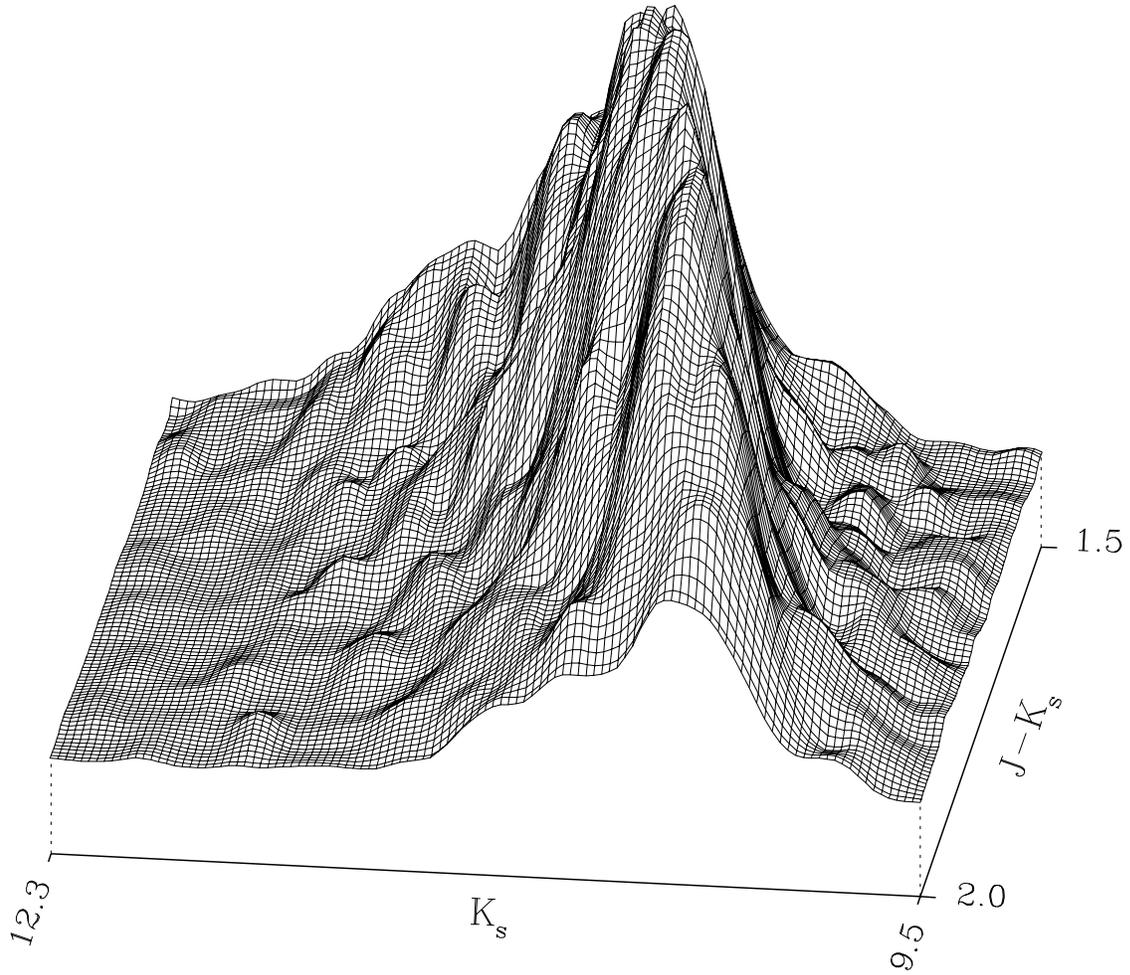}}
\caption{Density map of CMD Region~J.  The surface map covers color
range from $J-K_s=1.5$ to $J-K_s=2.0$.  Note the well-defined 
ridge, which suggests a luminosity-color relation for these stars.
\label{fig4}}
\end{figure*}

\begin{figure}[t!]
\epsfysize=11.0cm
\centerline{\epsfbox{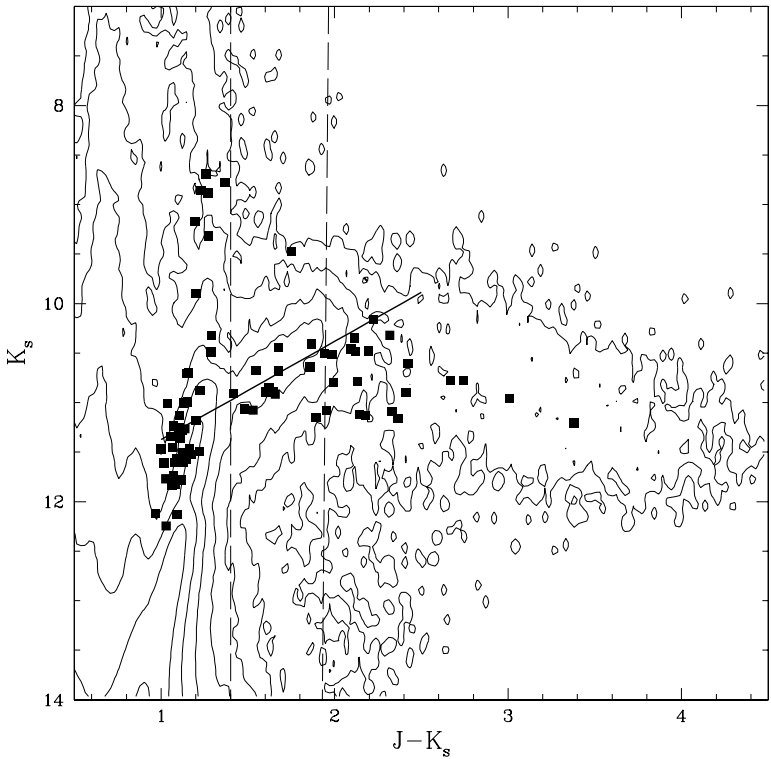}}
\caption{The sample of 79 oxygen- and carbon-rich Miras from Glass
\etal{} (1990), plotted over the color-magnitude diagram of 2MASS.
Magnitude and color of each Mira are converted to CIT photometric 
system and averaged over the period.  Vertical dashed lines show 
the color range, $1.4 < J-K_s < 1.9$, used for the straight line 
fit to luminosity-color relation. The best-fit LC relation is 
indicated with thick solid line.
\label{fig5}}
\end{figure}

The 2MASS sample of C-rich LPVs in the LMC contains 8229 stars.  The
surface map of Region~J is shown in Figure~\ref{fig4}.
The luminosity-color relation for these stars results in the
well-defined ridge in the figure.  In the absence of the period data,
we cannot use standard PL or PC relations to calibrate the intrinsic
brightness of these variables.  Rather, we have to rely on
luminosity-color relation.  Figure~\ref{fig5} shows the sample of 79
oxygen- and carbon-rich Miras in the LMC \cite{gla90}.  Magnitude and
color of each Mira are averaged over period and plotted on top of
2MASS color-magnitude digram.  The luminosity-color relation (shown
with the solid line) for 14 carbon Miras in the color interval bounded
by vertical dashed lines is
\[
<K_s> = (-0.99\pm0.80) <J-K_s> + (12.36\pm1.33), \qquad \sigma=0.38.
\]
The average r.m.s is $\sigma \simlt 0.3^m$ for $1.4 < J-K_s < 1.7$.
Given this LC relation, we may argue that selecting LPVs from a 
reasonably narrow color range will result in sources with similar 
luminosities, i.e., standard candles.  For our analysis, we choose 
color interval $1.6<J-K_s<1.7$, sufficiently narrow to ensure 
similar luminosities of our standard candles and sufficiently 
broad to host 
enough sources (1385) for statistically meaningful inference.

The ridge-line fit above was based on the average LC relation.  From
analysis of light curves in Glass \etal{} (1990), the amplitudes of
carbon-rich LPVs are $\Delta K \simlt 0.5$ mag, so one may expect
significant broadening of the LC relation due to random phase
observations.  However, as we will demonstrate below
(\S\ref{sec:sca_results}), the width of main peak in the apparent
luminosity function is only $\sigma\approx0.2^m$, even with random
phase observations.  This suggests that the phase-average LC relation
is much tighter.

\subsection{Method} \label{sec:sca_method}

\begin{figure}[h!]
\epsfysize=9.5cm
\centerline{\epsfbox{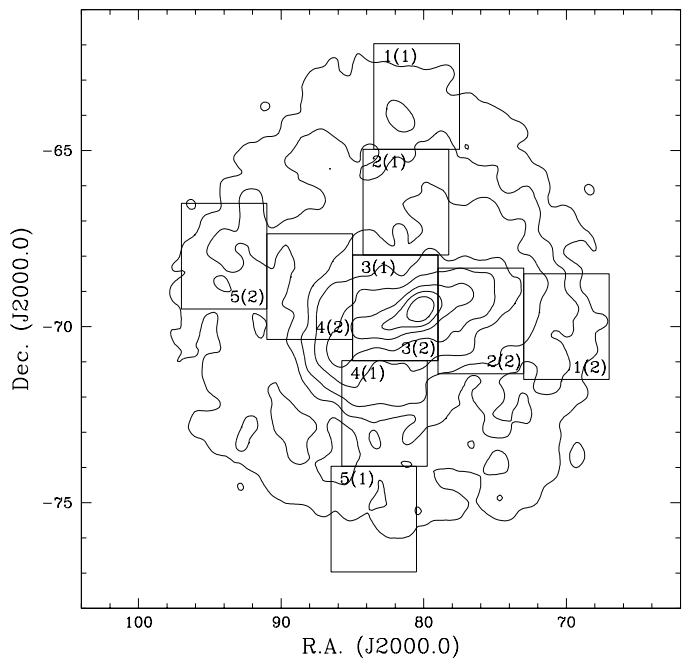}}
\caption{Locations of selected LMC fields.  Field positions follow 
  two arcs: Arc~1 (North-South direction, parallel to the line of 
  nodes), and Arc~2 (East-West direction, perpendicular to the 
  line of nodes).  Field numbers are indicated, with arc numbers 
  given in parentheses.  Field~3 is the same for both arcs.
\label{fig6}}
\end{figure}

Without prior knowledge of the true source distribution, which is
likely to be irregular due to tidal interaction (e.g. Weinberg 2000)
or sufficient characterization of the stellar populations to allow a
non-parametric density estimation with all of the data, we study the
photometric distribution in several fields in the LMC.  The fields are
located along two great circle arcs passing through the central region
of the Cloud: Arc~1, parallel to the line of nodes, and Arc~2,
perpendicular to the line of nodes (see Fig.~\ref{fig6}).

The apparent luminosity function can be analyzed in terms of the 
centroid $\overline m$ and the width $2\sigma_m$ of the 
distribution.  For a homogeneous standard candle population, the 
centroid of the distribution measures LMC distance and the width 
of the distribution gives an estimate of its line-of-sight depth.
The error analysis of a standard apparent magnitude-absolute 
magnitude-distance relation gives
\begin{equation}
\sigma_m^2\approx\sigma_M^2+\frac{4.72}{R_{LMC}^2}\sigma_r^2+
\sigma_A^2+\sigma_{ph}^2,
\label{eq:eqerr}
\end{equation}
where $R_{LMC} = 50$ kpc is the average distance to the LMC,
$\sigma_M$ is the intrinsic precision of our standard candles,
$\sigma_A$ is the variance due to extinction, $\sigma_r$ is the
geometric depth, and $\sigma_{ph}$ is the photometric error.
Equation~(\ref{eq:eqerr}) states that the apparent brightness
distribution is the convolution of the spatial density and the
intrinsic luminosity function and therefore provides an {\em upper
  limit} to the geometrical depth.

\subsection{Results and Interpretation} \label{sec:sca_results}

\begin{figure}[h!]
\epsfysize=11.0cm
\centerline{
\epsfbox{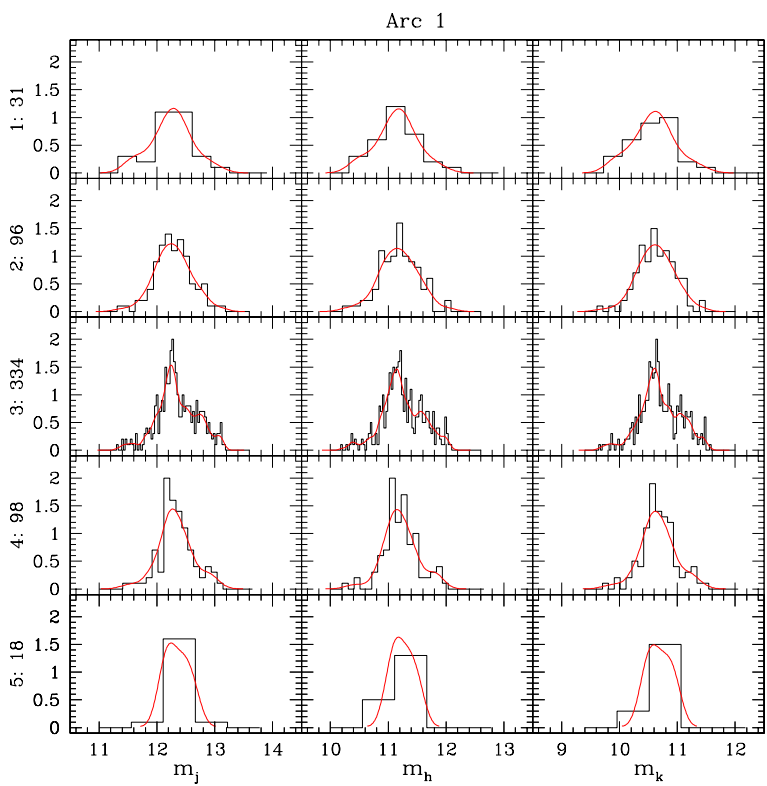}}
\caption{Apparent brightness distribution of selected C-rich Miras
  in fields along Arc~1.  Histograms are normalized raw data,
  smooth curves are kernel smoothed densities.  The bin widths in 
  each histogram are chosen to ensure signal-to-noise ratio of 3 
  or better.  Columns correspond to $J$, $H$, $K_s$ bands, rows 
  correspond to fields.  Fields and the numbers of C-rich LPVs in 
  the fields are labeled.  See Figure~\ref{fig6} for field 
  designations.
\label{fig7a}}
\end{figure}

\begin{figure}[h!]
\epsfysize=11.0cm
\centerline{
\epsfbox{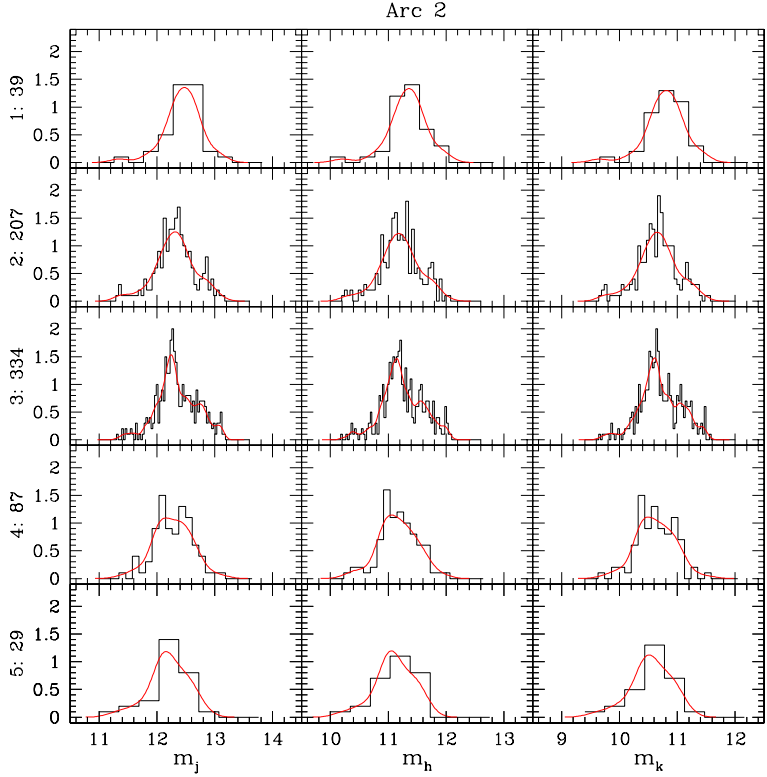}}
\caption{Same as Figure~\ref{fig7a}, but for fields along Arc~2.
\label{fig7b}}
\end{figure}

The brightness distributions of standard candles in selected 
fields are shown in Figures~\ref{fig7a} and \ref{fig7b}. 
All fields in Figures~\ref{fig7a} and \ref{fig7b} exhibit well-defined
central peaks corresponding to the midplane of the LMC disk.  We
immediately notice that the narrowest features in the brightness
distributions have widths $\sigma_m \approx 0.2^m$ (cf. Field~3).
This is a {\em direct evidence} that our color-selected sources are
standard candles at least as good as $\sigma_M\approx0.2^m$.  In fact,
they are even better, because of the additional terms on the
right-hand-side in equation~(\ref{eq:eqerr}).  This suggests that
carbon LPVs in the narrow ($\sim 0.1$ mag) color range are excellent
standard candles, even observed at random phases.

\subsubsection{Analysis of the Distribution Centroids}

\begin{figure}[h!]
\epsfysize=10.0cm
\centerline{
\epsfbox{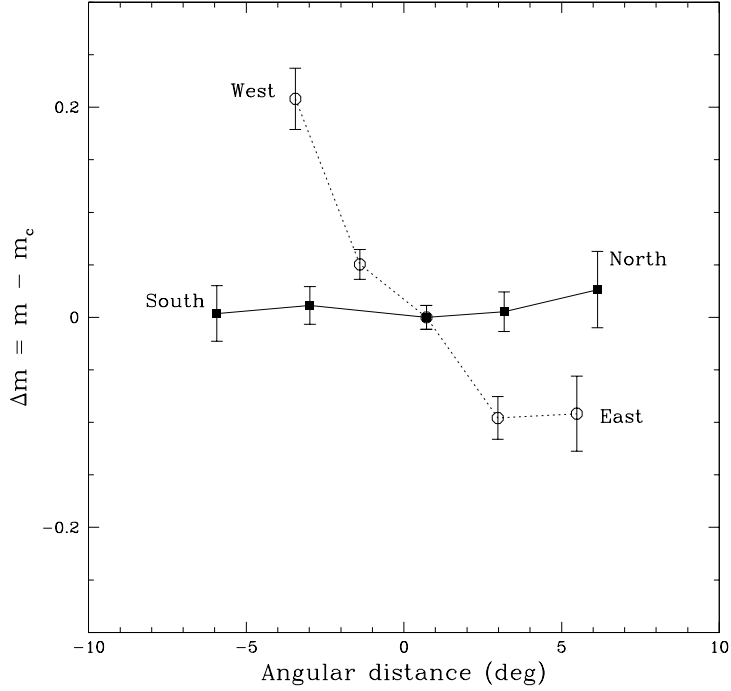}}
\caption{The mean magnitude offset averaged over $J$, $H$, $K_s$ 
  bands as the function of the angular distance from the optical 
  center of the bar.  The offsets are given relative to the central
  fields of the corresponding arcs.
\label{fig8}}
\end{figure}

The centroids of the distributions are consistent with the inclination
of the LMC derived in \S\ref{sec:ml}.  Since Arc~1 is parallel to the
line of nodes, the stars in the fields along this arc should be
roughly at the same distance.  Hence, we do not expect any drift in
the means $\overline m$ for stars in these fields.  On the other hand,
we expect a shift in the mean magnitude for fields along Arc~2, which
is perpendicular to the line of nodes.  Assuming the Eastern part of
the Cloud is closer to us, sources in Eastern fields should be, on
average, brighter than their counterparts in Western fields.
Figures~\ref{fig7a} and \ref{fig7b} confirm the expectations.  The
magnitude of the effect is shown in Figure~\ref{fig8} for both arcs as
the function of the angular distance from the optical center of the
bar.
To improve the signal-to-noise ratio,
we take the average $\overline m$ of the mean magnitudes in three
bands and plot the resulting averages normalized to the central 
field of the respective arc.  We find that fields along the 
North-South line have similar mean magnitude, whereas Eastern
fields are on average $0.1^m$ brighter and Western fields are 
$\sim 0.2^m$ fainter than central bar fields.  The magnitude 
difference is similar to what Caldwell~\& Coulson~(1986) found in
their analysis of Cepheids in the Cloud.

Data shown in Figure~\ref{fig8} allow independent determination 
of the LMC inclination angle.  We provide results of two methods:
(1) the inclination of the best fit plane to the photometric 
distance estimate for each field; and (2) the mean inclination of
each field paired with the central field.  In Method 1, we 
compute the best fit line through spatial positions of the field 
centers.  The estimates for each arc are $0.9^\circ\pm0.3^\circ$ 
and $42.3^\circ\pm7.2^\circ$, respectively.  We observe that the 
inclinations along our Arc~1, as expected, are consistent with 
zero, while the inclination angles along Arc~2 are consistent 
with values in Table~\ref{table:expd} at $2\sigma$ level.  
Alternatively, in Method 2 does not presuppose a tilted plane but
rather computes the inclination of each field relative to the 
line of sight through the central field.  We then compute the 
weighted average over all pairs.  The resulting inclinations are 
$1.0^\circ\pm4.1^\circ$ and $34.4^\circ\pm2.7^\circ$ for the 
first and second arcs, respectively.

In principle, this magnitude drift could have been produced by a
reddening gradient across the Cloud. However, the extinction maps
by Oestreicher~\& Schmidt-Kaler (1996) do not show any systematic
change in reddening in the West-East direction across the LMC. 
In addition, an analysis of the extinction across the entire 
cloud based on the location of the giant branch reveals little 
evidence for significant extinction on average outside of the 
inner square degree \cite{nik00}.  Moreover, if the magnitude drift
were indeed due to reddening gradient, it would have had 
band-dependent signature.  The ratios of magnitude offset in $J$, 
$H$, $K_s$ bands would be in direct proportion to the extinction 
coefficients $A_J$, $A_H$, $A_K$.  Since this is not the case, we 
have to reject this possibility and interpret the drift as
the true distance effect.

\subsubsection{Shape of the Distributions}

Careful study of the density distributions in Figures~\ref{fig7a}
and~\ref{fig7b} shows that some distributions have extended tails.
For example, distributions in Fields 3 and 4 clearly show positive
skewness.  Figure~\ref{fig4} also shows the extension of the central
ridge toward fainter magnitudes.  To quantify the extended component,
we model the distribution of LMC disk sources with a properly centered
Gaussian.  After subtracting the disk component in the central field
(Field~3), we find a secondary, broader distribution which contains
approximately 90 sources or about a quarter of the total number in the
field.  This distribution itself appears to have two distinct
components.  The strongest peak of this extended component is $0.5$
mag fainter than that of the disk distribution and a weaker peak is
roughly $0.8$ mag brighter.  In general, the tail toward fainter
magnitudes is more pronounced, although the distribution extends to
both sides.

Among the factors which could produce the extended component visible
in Figures~\ref{fig7a} and~\ref{fig7b} are: 1) foreground population,
2) source blending, 3) interstellar reddening, 4) population of
overtone pulsators, 5) distribution of periods, 6) age/metallicity
variations, and 7) spatial density distribution\footnote{We do not
  list spurious detection as the possibility, because the distribution
  s are the same in three bands, suggesting real feature}.  Foreground
population can be rejected because the fraction of Galactic sources in
Region~J is nearly zero (see Table~\ref{regions}).  Interstellar
reddening would produce a band-dependent effect (i.e. longer tails in
$J$ band than in $K_s$ band).  We consider each of the remaining five
possibilities in detail:

\begin{enumerate}
%
%
\item {\em Source blending}.  Considering that the secondary peak runs
  parallel to the main LC relation (cf.  Figure~\ref{fig4}), at
  least one member in the blend must be a carbon star.  However,
  blending a carbon star with an unresolved source would produce a
  source brighter than the carbon star.  This means the secondary peak
  would be brighter than the main feature, which contradicts to data;
%
%
\item {\em Population of overtone pulsators}. The question about the
  pulsation mode of Mira variables seem to have been resolved recently
  with Miras unambiguously identified as fundamental mode pulsators
  \cite{woo96,woo98}.  It is conceivable, then, that our
  color-selected sample of standard candles includes a population of
  first- and higher-overtone LPVs.  Based on Figures~\ref{fig7a}
  and~\ref{fig7b}, this `secondary' population should constitute about
  $25\%$ of the entire sample, and be fainter by $0.4-0.5^m$.  We test
  this possibility by cross-correlating 2MASS data with the sample of
  Hughes \& Wood (1990), which contains fundamental-mode LPVs only.
  Picking out sources in the color range $1.6<J-K_s<1.7$ and plotting
  the histogram of their $K_s$ magnitudes reveals the same extended
  component, approximately $0.5$ mag fainter than the main peak.
  Since the feature is present even in the data without overtone
  pulsators, this explanation has to be rejected;
%
%
\item {\em Distribution of periods}.  Even in case of zero-dispersion
  PLC relation for C-rich LPVs, the observed magnitude
  distribution could result from an intrinsic period distribution (in
  addition to overtone pulsations discussed above).  Although this
  explanation remains a possibility because the physics of these stars
  remains uncertain in detail, such an effect is not apparent in
  current theoretical models, which confirms an intrinsic width of the
  instability strip but not multiple peaks.  Conversely, as previously
  mentioned, the good fit of the primary peak to the disk inclination
  gives us confidence in a well-defined instability strip and width of
  this peak is consistent with other pulsators.
%
%
  Alternatively, the pulsation periods depend on mass and metallicity
  and distribution of periods implies the range of mass, ages and
  metallicities, and vice versa.  Assuming a range of initial AGB
  masses leads to the range of intrinsic bolometric magnitudes for a
  single period \cite{mar96}.  Marigo's \etal{} theoretical
  evolutionary tracks show the difference in a few tenths of a
  magnitude for carbon Miras at a constant period, depending on the
  mass of the star.  Fitting the apparent luminosity function to these
  tracks \cite{col00} suggests that the stronger peak may be due to
  younger ($\sim 0.6$ Gyr) and more massive stars ($2 M_\odot$), while
  the broader component is formed by older ($\sim 2.8$ Gyr) and less
  massive ($\sim 1.2 M_\odot$) stars.  This appears consistent with
  observations: Frogel \etal{} (1990) found that carbon stars are
  present in globular clusters aged 100 Myr to about 3 Gyr and that
  carbon stars in intermediate-age LMC clusters are a few tenths of a
  magnitude fainter than those in young LMC clusters.  Alves \etal{}
  (1998) reported that AGB variables in the intermediate age cluster
  NGC 1783 are about 0.5 mag brighter at given period than variables
  in ancient LMC globular cluster NGC 1898.  We find two aspects of
  this possibility to be unsatisfying.  First, it does not naturally
  explain the bright secondary peak. Second, the star formation rate
  subsequent to the 2.8 Gyr burst would have to be small in order to
  result in a well-defined feature;
%
%
\item {\em Extended density distribution in the LMC}.  This appears to
  be the most natural explanation to the observed brightness
  distributions: it is band-independent, it does not rely on special
  star formation history, and straightforwardly explains both the
  brighter and fainter secondary distribution.  The distinct feature
  at fainter magnitudes may be associated with the population
  kinematically distinct from the disk sources, recently found by
  Graff \etal {} (1999).  This motivates a detailed kinematic
  follow-up of the small bright ``clump'' in the photometric
  distribution. Based on the offset, this population is 15 kpc from
  the LMC center and coincident with the intervening population
  detected by Zaritsky \& Lin (1997).  The existence of LPVs implies a
  relatively young population, and the broad area is consistent with
  tidally stripped material.  A future detailed study will attempt to
  self-consistently model the disk, an extended spheroid component as
  suggested by Hughes~\etal{} (1991), and test for other distinct
  features.  Regardless of details, the broad photometric distribution
  suggests that LMC is {\em geometrically thick} along the line of
  sight.  From Figures~\ref{fig7a} and~\ref{fig7b}, the average
  thickness of the LMC is $2 \sigma_m \sim 0.7^m$, which implies
  a thickness (after deconvolution) of $\sim 14$ kpc (for the LMC
  distance $R_{LMC} = 50$ kpc).  This makes the LMC as thick along the
  line of sight as it is wide across the sky, something which one
  would naively expect for a dwarf companion in the process of being
  tidally shredded.  In summary, the spatial interpretation of these
  photometric distributions suggests the LMC consists of a centrally
  concentrated barred disk and some number of extended distributions,
  including a spheroid and tidally distorted and possibly stripped
  populations.
\end{enumerate}

\section{Summary} \label{sec:summary}

We have analyzed the spatial distributions of several LMC 
populations, identified in NW00 based on their location in the 
color-magnitude diagram.  Quantitative analysis of the observed 
distributions includes parametric source-count and a preliminary 
standard candle analyses.  Our major conclusions are:
\begin{itemize}
  
\item Projected star count analyses based 2MASS data yield scale
  lengths, deprojection-based inclinations, and position angles
  consistent with previous studies.  The near-far degeneracy of 
  the disk orientation is broken by perspective difference.  The 
  near side of the Cloud subtends a larger angle in the sky-and 
  this allows us to determine that Eastern side of the disk is 
  closer, in agreement with Cepheid-based results.
  
\item We propose using carbon-rich LPVs in a narrow color range, 
  $1.6 < J-K_s < 1.7$ as standard candles to probe the structure
  of the LMC along the line of sight.  Based on published light
  curves, their intrinsic magnitudes have a dispersion of $\simlt
  0.3^m$ including the random phase of the observations.  The 
  width of the LMC disk in the observed photometric distribution 
  ($\sigma_m \approx 0.2$) suggests that these are excellent 
  standard candles.
  
\item The photometric distribution of our standard candle sample reveals strong
  central peak with extended tails in both directions, with tail
  toward fainter magnitudes more pronounced.  In the densest field,
  the distributions in all bands show two secondary peaks, $0.5$ mag
  fainter and $0.8$ mag brighter than the primary.  We interpret the
  primary peak as due to the midplane of the LMC disk and examine
  various possibilities which could produce the tails and secondary
  peaks of observed distribution.  We conclude that the photometric
  distribution of standard candles is most likely caused by a spatially
  extended stellar component and tidally stripped debris, consistent
  with a tidally disturbed dwarf companion.  It is possible that the
  distribution reflects the intrinsic period distribution of LPVs but
  or distinct populations of masses (ages) and metallicities for these
  stars, but such explanations special populations or evolutionary
  histories.
  
\item The distribution of apparent magnitudes of standard candles is
  consistent with tilted geometry derived from star counts.  We derive
  a {\it direct} determination of the LMC disk inclination of
  $42.3^\circ\pm7.2^\circ$, consistent with Laney \& Stobie (1986)
  estimate of $45^\circ\pm7^\circ$ and with results of Welch \etal{}
  (1987), $37^\circ\pm16^\circ$.  Interpreting the apparent luminosity
  function as due to real source density, we find evidence of the
  extended component of the LMC, with a width of approximately 14~kpc
  (for the LMC distance $R_{LMC}=50$~kpc).
\end{itemize}

This detection of this thick component implies that LMC may contain a
kinematically distinct population as suggested by Graff \etal{} (1999)
and/or an the extended component found by Hughes~\etal{} (1991) and
predicted for the tidal interaction with the Milky Way (Weinberg
2000).  The existence of such a population will affect the
self-lensing models of the LMC in the microlensing studies.  These
preliminary results suggest a number of projects for short- and
long-term follow-up.  An improved analysis of the 2MASS sample may be
obtained with larger sample of standard candles, or with full three
dimensional analysis based on several distinct standard candles using
data from the entire survey rather than selected fields (work in
progress).  Assuming that the extended stars originated in the disk,
both populations will appear rotationally supported.  However, using
the 2MASS photometric distribution as a guide, a combined disk and
extended spheroid population should be kinematically separable and
these stars are good candidates for future spectroscopic work.

\section*{Acknowledgements}
The authors would like to thank Peter Wood for MACHO data on 
variable stars and for help in interpretation of these data.
We are grateful to Michael Skrutskie and Andrew Cole for
careful reading of the manuscript and suggesting ways of 
improving it.  We thank Shashi Kanbur for helpful advice regarding 
LPVs.  This publication makes use of data products from the Two 
Micron All Sky Survey, which is a joint project of the University 
of Massachusetts and the Infrared Processing and Analysis Center, 
funded by the National Aeronautics and Space Administration and 
the National Science Foundation.


\begin{onecolumn}
\begin{appendix}

\section{Parametric Maximum Likelihood} \label{sec:appdetails}

The expected source density for a bin in the direction 
($\alpha_i$, $\delta_j$) is determined by integrating the LMC 
density model along the line of sight across the bin.  This is 
given by the following integral:
\begin{eqnarray} \label{eq:nexp}
\rho^{exp}_{ij} &\propto& \int _0 ^\infty dt\, t^2
\int _{\delta_j - \Delta \delta/2} ^{\delta_j + \Delta \delta/2} 
d\delta \cos \delta
\int _{\alpha_i - \Delta \alpha/2} ^{\alpha_i + \Delta \alpha/2} 
d\alpha\, \rho (t, \alpha, \delta) \nonumber \\
&\approx&
\int _0 ^\infty dt\, t^2 \rho (t, \alpha_i, \delta_j) 
\cos \delta_j\, \Delta \alpha\, \Delta \delta,
\end{eqnarray}
where the last equality assumes that the bin size $\Delta
\alpha$, $\Delta \delta$ is sufficiently small.  We perform the
integral in (\ref{eq:nexp}) using 256-point Gaussian quadrature
formula, with $20$~kpc and $80$~kpc as the integration limits.  
The underlying source density $\rho (t, \alpha, \delta)$ is given
either by equation~(\ref{eq:expd}) or by equation~(\ref{eq:powl}).
The coordinate transformations for both models follow.

\subsection{Exponential Disk} \label{sec:appexpd}

To quantify $\rho (\cdot)$, we introduce the coordinate system 
$\{ x_0, y_0, z_0 \}$ which has the origin at the center of the 
LMC at $\{ t, \alpha, \delta \} = \{ R_{LMC}, \alpha_0, \delta_0 
\}$ and has $z_0$-axis toward the observer, $x_0$-axis antiparallel 
to the right ascension axis, and $y_0$-axis parallel to the 
declination axis.  The coordinate transformations are given by
\begin{eqnarray} \label{eq:x0y0z0}
x_0 &=& -t \cos \delta \sin (\alpha-\alpha_0) \nonumber \\
y_0 &=& t \sin \delta \cos \delta_0 - t \cos \delta 
\sin \delta_0 \cos (\alpha-\alpha_0) \\
z_0 &=& R_{LMC} - t \cos \delta \cos \delta_0 
\cos (\alpha-\alpha_0) - t \sin \delta \sin \delta_0. \nonumber
\end{eqnarray}
The coordinate system of the exponential disk, $\{ x', y', z' \}$,
is the same rectangular system as $\{ x_0, y_0, z_0 \}$, except 
rotated about $z_0$-axis by the position angle $\theta$ 
counterclockwise and about the new $x'$-axis by inclination 
angle $i$ clockwise.  The coordinate transformations are given by
\begin{eqnarray} \label{eq:xpypzp}
x' &=&  x_0 \cos \theta + y_0 \sin \theta \nonumber \\
y' &=& -x_0 \sin \theta \cos i + y_0 \cos \theta \cos i - 
z_0 \sin i \\
z' &=& -x_0 \sin \theta \sin i + y_0 \cos \theta \sin i + 
z_0 \cos i.  \nonumber
\end{eqnarray}
Because our resolution in photometric distance is larger than the
disk thickness, the integral in equation~(\ref{eq:nexp}) may be 
simplified by assuming that the exponential disk is infinitely 
thin, i.e. $z' = 0$ for all points of the disk.  The contribution 
to the integral is zero everywhere, then, except the point where 
line of sight intercepts the plane of the disk.  The value of $t$ at
the intercept, $\overline t$, is
\begin{eqnarray} \label{eq:t}
\overline t &=& -R_{LMC} \cos i \times \left[ \cos \delta 
\sin (\alpha-\alpha_0) \sin \theta \sin i\right. \nonumber \\
&&\left. +
\left( \sin \delta \cos \delta_0 - \cos \delta \sin \delta_0 
\cos (\alpha-\alpha_0) \right) \cos \theta \sin i \right.\\
&&\left. -
\left( \cos \delta \cos \delta_0 \cos (\alpha-\alpha_0) + 
\sin \delta \sin \delta_0 \right) \cos i \right]^{-1}.
\end{eqnarray}
The values of $x_0$, $y_0$ and $z_0$ coordinates at the 
intercept point follow from equation~(\ref{eq:x0y0z0}).  The
radius of the intercept point in the plane of the exponential disk
is given by
\begin{equation} \label{eq:r}
r = \sqrt {x'^2 + y'^2}.
\end{equation}
Finally, the expected source 
density~(\ref{eq:nexp}) may be written as
\begin{equation}
\rho^{exp}_{ij} \propto \overline t^2 {\rm exp} (-r/R) \cos \delta_j,
\end{equation}
where $\overline t$ and $r$ are given by equations~(\ref{eq:t}) 
and~(\ref{eq:r}), respectively.

\subsection{Power Law Model} \label{sec:apppowl}

Treatment of the spherical power law model is much simpler. Unlike
the disk, there is no unique axis of symmetry and one may write 
the density~(\ref{eq:powl}) in the coordinates $t$, $\alpha$ and 
$\delta$ directly. It is straightforward to verify that
\begin{equation}
r = \sqrt {t^2 + R_{LMC}^2 - 2R_{LMC}t
\left[ \cos \delta \cos \delta_0 \cos (\alpha-\alpha_0) + 
\sin \delta \sin \delta_0 \right]}.
\end{equation}

\end{appendix}
\end{onecolumn}

\eject

\end{document}